\documentclass[iicol]{sn-jnl}

\usepackage{graphicx}%
\usepackage{multirow}%
\usepackage{amsmath,amssymb,amsfonts}%
\usepackage{amsthm}%
\usepackage{mathrsfs}%
\usepackage[title]{appendix}%
\usepackage{xcolor}%
\usepackage{textcomp}%
\usepackage{manyfoot}%
\usepackage{booktabs}%
\usepackage{algorithm}%
\usepackage{algorithmicx}%
\usepackage{algpseudocode}%
\usepackage{listings}%
\usepackage{dsfont}
\usepackage{color}
\usepackage{longtable}
\usepackage{verbatim} 

\begin{document}

\title[Resonant wavelengths of whispering gallery modes in dispersive materials]{Resonant wavelengths of whispering gallery modes in dispersive materials}                               
\author*[1]{\fnm{Lorena} \sur{Velazquez-Ibarra}} 
\email{lorenav@fisica.ugto.mx}

\author[1]{\fnm{Juan} \sur{Barranco}}
\email{jbarranc@fisica.ugto.mx}

\affil*[1]{\orgdiv{Divisi\'on de Ciencias e Ingenier\'ias},  \orgname{Universidad de Guanajuato, Campus Le\'on},
 \postcode{37150}, \city{Le\'on}, \state{Guanajuato}, \country{M\'exico}}


\abstract
{In this work we compute the resonant wavelengths of whispering gallery modes for bulk-fused silica microspheres taking into consideration the material chromatic dispersion. This is done by following two approaches: solving the exact characteristic equation and solving the nonlinear equations that result from the asymptotic approximations for a wavelength-dependent refractive index. Similar results with both methods are obtained with differences below $1\%$. Nevertheless, important differences are found between the resonant wavelengths computed for a non-dispersive and a dispersive material. We compute the free spectral range and the quality factor, and make a comparison between the variable index and the constant index cases. Our results show that the quality factor is unaffected by chromatic dispersion. However, there are differences of significant relevance in the case of the free spectral range. Our work could be useful as a pathway for designing microspheres for different applications.}




\maketitle

\section{Introduction}

Whispering gallery modes (WGM) are waves that travel along the surface of a resonator with circular symmetry.  They where studied for the first time for acoustic waves by Lord Rayleigh in St. Paul's Cathedral~\cite{Rayleigh:1910}. Electromagnetic WGM in optical frequencies were first observed by Mie in scattering experiments~\cite{Mie:1908}, and have since been demonstrated for different resonator morphologies such as: spheres, cylinders, rings, toroids and disks~\cite{Vahala:03,Chiasera:10,Lin:19}. 
Braginsky et al.~\cite{Braginsky:89} successfully fabricated in 1989 solid glass microspheres that exhibit WGM with a quality factor ($Q$) as high as $10^8$ and, since then, a huge number of applications have been developed: from add-drop devices~\cite{Cai:99} and biosensors based on proteins attached to the surface of microspheres and microcylinders~\cite{Arnold:03,Hsieh:22,Zhou:23}, to light sources based on nonlinear and quantum optics~\cite{Strekalov:16,Huet:16,Soltani:18,Xavier:21}. It is noteworthy to mention their applications as microlasers~\cite{Chiasera:10} or as enhancers of photocell plates that increase their energy generation~\cite{Grandidier:11,Yao:12}.

Given the vast range of applications, the design of the microresonator (which includes geometry~\cite{Buck:03}, material and desired wavelength of operation) is an important task that involves the computation of the resonant wavelengths. 

Every material exhibits a degree of chromatic dispersion, i.e., different wavelength components travel at different velocities in the material, which is characterized by a wavelength-dependent refractive index.  This material dispersion plays an important role in the total dispersion of the resonator and should be taken into account in the computation of the resonant wavelengths.

In this work, we focus on WGM bulk-fused silica spherical microresonators and investigate the effect of the material chromatic dispersion in the calculation of the resonance positions. The analysis can be extended to cylindrical systems in a straightforward manner. For other geometries or non-dielectric materials, such as metals, different approaches to the one presented here should be adopted \cite{Demchenko:13,Lobanov:19,Yan:20}. 

The resonant wavelengths are solutions of an eigenvalue equation that involves transcendental functions which can be solved either through numerical methods or analytical asymptotic approximations~\cite{Lam:92,Schiller:93}. In these references, the accuracy of the asymptotic formulas was estimated under the assumption of a non-dispersive material, where the refractive index remains constant. However, when dealing with dispersive media, where the refractive index of the microsphere is wavelength-dependent (i.e. $n_s=n_s(\lambda)$), the accuracy of the asymptotic formulas is not evident since the equations become nonlinear in $\lambda$. The use of the asymptotic approximations is common practice for non-dispersive media \cite{Zhou:23,Zhou:24}, but also in the context of dispersive media ~\cite{Liang:15,Lin:16,Soltani:18}, despite the fact that the convergence of the asymptotic expansion has not been previously explored for the dispersive case. Here, we investigate the accuracy of the asymptotic formulas for a dispersive material and demonstrate their validity. As anticipated, we observe that the accuracy of the formulas improves with increasing azimuthal number, $l$. Our findings reveal that the relative error does not change significantly across different values of the radius of the sphere, $R$, and radial modes, $n$, and that it is of the order of a few percent. While this error can, in general, be considered small, it could become relevant, for instance, in sensing applications~\cite{Hanumegowda:05,Foreman:15}. 

Additionally, asymptotic formulas typically involve the computation of $n_sx$, where $x=2\pi/\lambda R$ is the resonance size parameter. When assuming a constant refractive index, $n_s$, the dependence of the resonant wavelength on the radius of the sphere is explicit once $x$ is computed. For dispersive media, however, the inverse relation between the resonance size parameter does not hold for different values of $R$ and, therefore, should be solved individually for each chosen value of $R$. Our analysis shows that the constant refractive index case should be taken with care.

In order to do this, the article is organized as follows: Section~\ref{section2} introduces the basic equations used to find the resonant wavelengths and the theoretical considerations/hypotheses under which the asymptotic formulas are valid. In Section~\ref{section3}, we present the computed resonant frequencies and compare the results obtained using the eigenvalue equation with the results of the asymptotic formulas in both non-dispersive and dispersive media. 
In order to quantify the effect of the dispersion, in Section~\ref{section4} we compute the free spectral range (FSR) and the $Q$-factor, and compare the results between the constant and wavelength-dependent refractive indices.
Lastly, in Section~\ref{conclusions} we conclude. 

\section{WGM Resonance Positions}\label{section2}

\subsection{Characteristic Equation}

The problem of propagation of electromagnetic fields in a dielectric sphere is well known and has been previously studied~\cite{Stratton, Kerker}, where solving the Helmholtz equation in spherical coordinates is simplified introducing vectorial spherical harmonics. The angular dependence of the electric and magnetic fields, $\vec E$ and $\vec B$, is thus completely determined, remaining a single equation for the radial part of the fields, $F_l(r)$. 
Demanding continuity of the tangential components of the fields at the surface of the sphere yields characteristic equations for two states of polarization: transverse electric (TE, where the electric field is parallel to the sphere surface) and transverse magnetic (TM, where the  magnetic field is parallel to the sphere surface) modes.
The resonant wavelengths are solutions of these equations and are characterized by a set of numbers $\{l,m,n,p\}$, where $n$ is the radial mode number, $l$ the polar mode number, $m$ the azimuthal mode number, and $p$ is related to the polarization mode. For each value of $l$, the azimuthal mode numbers can take values in the range $-l\leq m \leq l$, and the condition  $l=m$ describes the fundamental modes.

The characteristic equations for TE and TM modes in a microsphere of radius $R$ are~\cite{Chiasera:10}
\begin{equation}
    p(\lambda)\,\frac{J'_{\nu}(k(\lambda)\,n_s(\lambda)\,R)}{J_{\nu} (k(\lambda)\, n_s(\lambda)\,R)}=\frac{H'_{\nu}(k(\lambda)\, n_o(\lambda)\,R)}{H_{\nu} (k(\lambda)\, n_o(\lambda)\,R)},
    \label{eqcomplete}
\end{equation}
where 
\begin{equation}
     p(\lambda)=
    \begin{cases}
   n_{s}(\lambda)/n_{o}(\lambda) &\text{for TE modes}\\
     n_{o}(\lambda)/n_{s}(\lambda) &\text{for TM modes,}
\end{cases}
\label{TETM}
\end{equation}
$J_\nu$ and $H_\nu$ are the Bessel and Hankel functions of the first kind, respectively, of half integer order $\nu$, 
\begin{equation}
\nu~=l+\frac{1}{2}\,,
\end{equation} 
$k=2\pi/\lambda$ is the wave number, and $n_s$ and $n_o$ are the refractive indices of the sphere and the material outside the sphere, respectively, with the condition $n_s>n_o$ for confined modes. For the purposes of this work, we consider air as the material surrounding the resonator, so $n_o=1$.

For the following calculations, it is convenient to define the size parameter $x=k R$. Taking into account Snell's law, total internal reflection implies the fulfillment of the following condition for WGM resonances:

\begin{equation}
    x \le \nu \le  n_s x. \label{condition}
\end{equation}

From this point onwards, we drop the explicit dependence on $\lambda$ from the notation $p(\lambda),  k(\lambda)$, $n_s(\lambda)$, although it is implied unless otherwise specified.

In general, Eq.~\eqref{eqcomplete} is a complex equation for $\lambda$ (or, equivalently, for the frequency $\omega=2\pi c/\lambda $). The information of the resonant wavelengths is included in the real part of this equation, while the imaginary part contains information about intrinsic radiation losses. For the purposes of studying the validity of the asymptotic formulas for the resonant wavelengths, we focus the first part of our study on the real part of Eq.~\eqref{eqcomplete} which, for small values of the imaginary part of $\omega$, can be reduced to~\cite{AbraSteg72}
\begin{equation}
    p\,\frac{J'_{\nu}(n_s x)}{J_{\nu} (n_s x)}=\frac{Y'_{\nu}(x)}{Y_{\nu} (x)},
    \label{eqcar}
\end{equation}
where $Y_\nu$ is the Neumann function.

The solution of the characteristic equation for each polarization mode $p$ and for a fixed value $l=m$ consists on a series of discrete wavelengths, each corresponding to different values of the radial mode number $n$. The radial mode number is the number of maxima in the radial distribution of the field. The first radial mode, $n=1$, corresponds to the first root of Eq.~\eqref{eqcar}, before the strongly oscillating behavior of the Bessel functions begins. To find the position of the resonances $\{l,n,p\}$, Eq.~\eqref{eqcar} must then be solved numerically for $\lambda_{n,l}^{p}$.

\subsection{Asymptotic Approximation}

Lam et al.~\cite{Lam:92} derived an analytical expression for the resonances based on the asymptotic approximation of Eq.~\eqref{eqcar}. The derivation goes as follows:

 Classically, $\nu$ corresponds to the angular momentum which is related with the incidental angle as $\nu=~n_s x \sin(\theta_i)$. WGM are well confined if $\theta_i\sim \pi/2$ and, in consequence, $|n_sx -\nu|$ should be a small quantity for large values of $\nu$ since Eq.~\eqref{condition} must be satisfied for all $l$.
Therefore, it is possible to introduce a variable $t$ such that

\begin{equation}
\frac{n_s x-\nu}{\nu^{1/3}} = t\,, 
\label{expansion}
\end{equation}

\noindent{which, for large values of $l$, is expected to be a small parameter. Well known expressions for the Bessel and Neumann functions in terms of $n_s x=\nu+t \nu^{1/3}$ will be helpful in solving Eq.~\eqref{eqcar}. Indeed, it can be found that ~\cite{AbraSteg72,Chiasera:10}}

\begin{eqnarray}
J_\nu(\nu+t\nu^{1/3})&\sim& \left(\frac{2}{\nu}\right)^{1/3}Ai(-2^{1/3}t)\left[1+\sum_{j=1}^\infty\frac{f_j(t)}{\nu^{2j/2}}\right]\nonumber \\
&+&\frac{2^{2/3}}{\nu}Ai'(-2^{1/3}t)\sum_{j=0}^{\infty}\frac{g_j(t)}{\nu^{2j/3}}\,,\label{bessel}
\end{eqnarray}


and

\begin{eqnarray}
Y_\nu(\nu&+&t\nu^{1/3})\sim-\frac{e^{\nu(\cosh^{-1}(\nu/x)-\sqrt{1-(x/\nu)^2})}}{\sqrt{\pi\nu\sqrt{1-(x/\nu)^2}}}\nonumber\\
&\times&\left[1+\sum_{j=1}^\infty(-1)^j\frac{u_j((1-(x/\nu)^2)^{-1/2}}{\nu^j}\right]\,,\nonumber \\
\label{neumann}
\end{eqnarray}


\noindent{where $A_i$ is the Airy function and $f_j$, $g_i$ and $u_i$ are polynomials. Similar expansions for $J_\nu'(x)$ and $Y_\nu'(x)$  can be found elsewhere \cite{AbraSteg72}. Substituting those series in Eq.~\eqref{eqcar} and keeping only the first order term in the expansion, Eq.~\eqref{eqcar} reduces to}

\begin{equation}
-p\left(\frac{2}{\nu}\right)^{1/3}Ai'(-2^{1/3}t)=-\sqrt{\left(\frac{\nu}{x}\right)^2-1}Ai(-2^{1/3}t)\,. 
\label{eq_asympt}
\end{equation}

In the limit $\nu \to \infty$, the left hand side of Eq.~\eqref{eq_asympt} is zero, so that the equation can only be balanced if the argument $-2^{1/3}t$ is a zero of the Airy function. Hence, one can find a solution for $x$ and, consequently, the resonant wavelength, $\lambda_{n,l}^p$, can be computed as 

\begin{equation}
    \lambda_{n,l}^p=2\pi R n_s \left(\nu-2^{-1/3}t_n^0\nu^{1/3}+\mathcal{O}(\nu^{-1/3})\right)^{-1}\,,
    \label{orderzero}
\end{equation}
where $t_n^0$ is a zero of the Airy function. 
Further higher order corrections to Eq.~\eqref{orderzero} are obtained by adding new terms in powers of $\nu^{-n/3}$,
\begin{equation}
   t_n^0 =-2^{1/3}t+\sum_{q=1}^{\infty}c_q\nu^{-q/3}\,,
\end{equation}
so that an approximate solution~\cite{Lam:92, Datsyuk:92, Schiller:93} can finally be obtained:

\begin{equation}
\lambda_{n,l}^p=2\pi R n_s
\left(\nu - t^0_n \left(\frac{\nu}{2} \right)^{1/3} + \sum_{q=0}^N c_q\nu^{-q/3}\right)^{-1}\,,
\label{Eq:Lam}
\end{equation}
where the coefficients $c_q$ are given by

\begin{eqnarray}
c_0&=&-\frac{p}{\sqrt{n_s^2-1}}\,,\nonumber\\
c_1&=&\frac{3}{2^{2/3}10}(t_n^0)^2\,,\nonumber\\
c_2&=&\frac{1}{2^{1/3}}p \frac{t_n^0(1-\frac{2}{3}p^2)}{\left(n_s^2-1\right)^{3/2}}\,,\\ 
&\vdots& \nonumber
\label{coefLam}
\end{eqnarray}

So far, we have neglected the imaginary part of Eq.~\eqref{eqcomplete}.
The accuracy of the asymptotic expansions given by Eq.~\eqref{Eq:Lam} has been studied before~\cite{Lam:92,Schiller:93}, but under the assumption of a constant refractive index $n_s$. In the following sections, we test the accuracy of those expressions in the case of a dispersive material, taking into account the explicit dependence of the refractive index on wavelength, $n_s(\lambda)$.

\section{Exact vs Asymptotic}\label{section3}

\begin{figure}[h]
    \centering    
    \includegraphics[angle=0,width=0.47\textwidth]{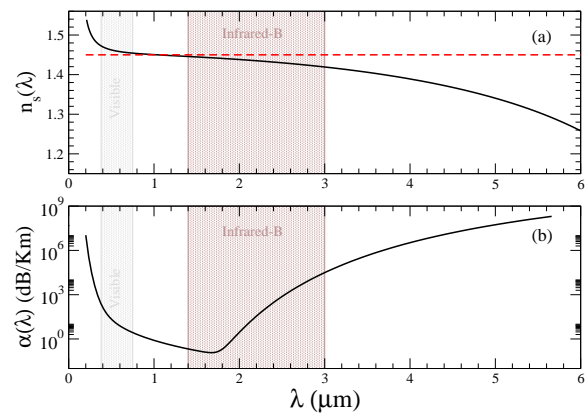}
    \caption{(a) Refractive index, $n_s$, and (b) absorption coefficient, $\alpha$, for bulk-fused silica as a function of wavelength, calculated using Eqs.~\eqref{sellmeier} and  \eqref{alpha}, respectively.}
    \label{fig:index}
\end{figure}

\begin{table}
\begin{tabular}{@{}cc@{}}
\toprule
$a_i$ & $b_i  (\mu$m) \\
\midrule
$a_1=0.6961663$ & $b_1=0.0684043$ \\ 
$a_2=0.4079426$   & $b_2=0.1162414$ \\  
$a_3=0.8974794$ &  $b_3=9.8961610$ \\ 
\botrule
\end{tabular}
\caption{Sellmeier coefficients for bulk-fused silica~\cite{Agrawal}.} \label{sellm-coef1}
\end{table}

\begin{table*}
\begin{tabular*}{\textwidth}{@{\extracolsep\fill}c cccc}
\toprule
 & \multicolumn{4}{@{}c@{}}{
        $\lambda_{1,l}^{TE} (\mu$m), $n_s(\lambda)$}    \\\cmidrule{2-5}
$l$     & $\lambda_{1,l}^{exact}$ & $\lambda_{1,l}^{asymp}, q=2$ & $\lambda_{1,l}^{asymp}, q=1$ &$\lambda_{1,l}^{asymp}, q=0$  \\ \midrule
$9$     & $3.637635$  & $3.641827$ & $3.532485$  & $3.666364$ \\ \hline
$13$    & $2.708779$ & $2.690939$ & $2.644735$ & $2.711827$ \\ \hline
$15$    & $2.398249$ & $2.383584$ & $2.350675$ & $2.401292$ \\ \hline
$20$    & $1.865201$ & $1.858348$  & $1.841854$ & $1.870123$\\ \hline
$30$    & $1.297424$ & $1.295927$  & $1.289815$ & $1.301912$\\ \hline
$50$   &  $0.813333$ & $0.813121$  & $0.811423$ & $0.815438$\\ \hline
$100$   &  $-$ & $0.427212$  & $0.426931$ & $0.427789$\\
\botrule
\end{tabular*}
\caption{TE Resonance positions calculated with different approximations, for a microsphere of radius $R=\,5\,\mu$m, radial mode $n=1$, and different azimuthal numbers, $l$ and considering chromatic dispersion, $n_s(\lambda)$.}
\label{lamres1}
\end{table*}

\begin{table*}
\begin{tabular*}{\textwidth}{@{\extracolsep\fill}c cccc}
\toprule
 & \multicolumn{4}{@{}c@{}}{
        $\lambda_{1,l}^{TE} (\mu$m), $n_s=1.45$ }  \\\cmidrule{2-5}%
$l$     &  $\lambda_{1,l}^{exact}$ & $\lambda_{1,l}^{asymp}, q=2$ & $\lambda_{1,l}^{asymp}, q=1$ &$\lambda_{1,l}^{asymp}, q=0$ \\ \midrule
$9$     & $3.756254$ & $3.741830$ & $3.633436$& $3.780535$\\ \hline
$13$    & $2.749168$&  $2.730287$ & $2.684063$ & $2.754484$\\ \hline
$15$    & $2.425057$ & $2.410426$ & $2.377460$ &  $2.43001$\\ \hline
$20$    & $1.876784$ & $1.870150$ & $1.853599$& $1.882516$\\ \hline
$30$     & $1.299994$ & $1.298513$ & $1.292365$& $1.304604$\\ \hline
$50$   & $0.811655$& $0.811438$& $0.809718$ & $0.813761$\\ \hline
$100$  & $0.414635$& $0.422203$ & $0.421908$ & $0.422778$\\
\botrule
\end{tabular*}
\caption{Same as Table~\ref{lamres1} but for a constant refractive index, $n_s=1.45$.}
\label{lamres2}
\end{table*}

In order to study the effect of chromatic dispersion, for definiteness in this work, we consider a bulk fused-silica microsphere and air as the exterior material. 
In particular, fabrication methods for silica micrometer spheres of different sizes are well established and the material optical properties are extensively characterized. Indeed, the material chromatic dispersion can be expressed through 
the Sellmeier equation~\cite{Agrawal},

\begin{equation}
n_s^2(\lambda)=1+\sum_{j=1}^3 \frac{a_j \lambda^2}{\lambda^2-b_j^2}\,,
\label{sellmeier}
\end{equation}
where the Sellmeier coefficients $a_j$ and $b_j$ for bulk-fused silica are given in Table~\ref{sellm-coef1}. 

Furthermore, the absorption coefficient, $\alpha$, has been measured and can be extrapolated as the following function~\cite{Gorodetsky:00}: 

\begin{eqnarray}
    \nonumber \alpha(\lambda)&=&\biggr(0.7\,\frac{\mu\mbox{m}^4}{\lambda^4}+1.1\times10^{-3}\exp{\frac{4.6\mu\mbox{m}}{\lambda}}+\\ 
    &&+4\times10^{12}\exp{\frac{-56\mu\mbox{m}}{\lambda}}\biggr)\frac{dB}{Km}\,. \label{alpha}
\end{eqnarray}

Fig.~\ref{fig:index} shows (a) the refractive index, $n_s$, and (b) the absorption coefficient, $\alpha$, as a function of wavelength for bulk-fused silica, calculated from Eqs.~\eqref{sellmeier} and \eqref{alpha}, respectively. 

The ``exact" resonant wavelengths for a fused-silica microsphere of radius $R=5\,\mu\mbox{m}$ are obtained as a numerical solution of Eq.~\eqref{eqcar} with $n_s$ given by Eq.~\eqref{sellmeier}. We have solved for the first radial order, $n=1$, and different values of the azimuthal number, $l$. The numerical solution was found with the secant method up to an accuracy of 16-digits. 
Identifying the different radial modes $n$ for large values of $l$ can be challenging due to the oscillatory behavior of the Bessel functions. Therefore, in order to ensure the correct identification of radial modes, it is critical that the numerical algorithm to find the roots of Eq.~\eqref{eqcar} have a reliable initial wavelength guess. In our case, we have used the wavelengths obtained with the asymptotic expansion~\eqref{Eq:Lam} as the initial seed.

Similarly, for the approximated solutions that use the asymptotic expansions for dispersive media $n(\lambda)$, we solved numerically Eq.~\eqref{Eq:Lam} for different orders in $\nu^{-q/3}$, namely $q=0,1,2$, using the coefficients given by Eqs.~\eqref{coefLam}, for the first radial order, $n=1$, and different values of the azimuthal number, $l$. 
In this case, Eq.~\eqref{Eq:Lam} has been solved using a damped Newton-Raphson method with an accuracy of 16-digits. 
We report the wavelength resonances $\lambda_{1,l}$ computed for TE modes with $n_s~=~n_s(\lambda)$ (see Table~\ref{lamres1}) and $n_s~=~1.45$ (see Table~\ref{lamres2}), for both the ``exact" and the asymptotic equations. 

\begin{figure}
    \centering    
    \includegraphics[angle=0,width=0.47\textwidth]{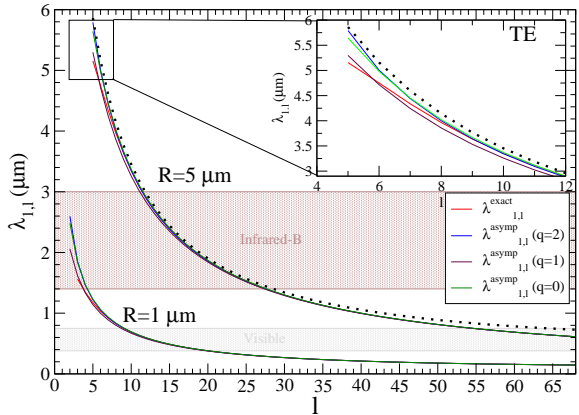}
    \caption{Dispersion curves for the first radial TE modes of $R~=1\,\mu$m and $R=5\,\mu$m microspheres, calculated with different approximations and $n_s(\lambda)$. The dotted line corresponds to the ``exact"  dispersion curve calculated for the $R=1\,\mu$m resonator and escalated for $R=5\,\mu$m.}
    \label{fig:frequencies}
\end{figure}

The values displayed in Table~\ref{lamres1} are plotted as dispersion curves (resonant wavelength as a function of the azimuthal number, $l$) in Fig.~\ref{fig:frequencies} for microresonators with $R=1\,\mu$m and $R=5\,\mu$m. As expected, the results are similar in all cases for $l \gg 1$. However, significant differences appear for small values of the azimuthal number $l$ (see inset of Fig.~\ref{fig:frequencies}). 

As previously mentioned, in the case of non-dispersive materials, the resonant wavelengths are computed directly from the size parameter $x$ and, for different sphere radii, these wavelengths are simply rescaled from $x$. In contrast, when taking into consideration the chromatic dispersion $n_s(\lambda)$, the resonant wavelengths are computed through a nonlinear equation and, consequently, can no longer be rescaled from $x$. This is illustrated in Fig.~\ref{fig:frequencies}, where we show the comparison between the resonant wavelengths computed for $R=5\,\mu$m (solid line) and for $R=1\,\mu$m rescaled to $R=5\,\mu$m (dotted line), and significant differences are evident. Finally, Fig.~\ref{fig:frequencies} also includes the results for different orders of approximation in the asymptotic formulas.

\begin{figure}
    \centering    
    \includegraphics[angle=0,width=0.47\textwidth]{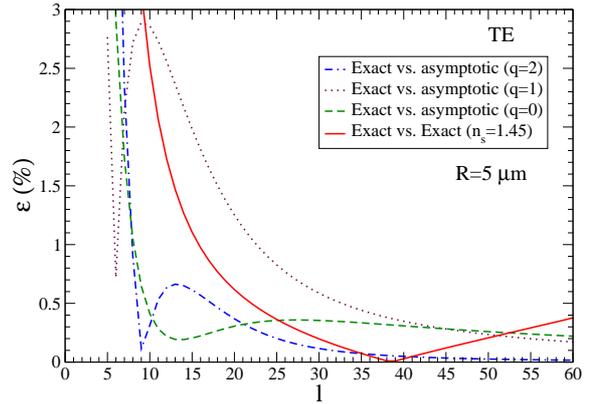}
    \caption{Percent relative error computed between the 
    ``exact" and the different asymptotic solutions for a silica microsphere with $R=5\,\mu$m, for TE modes, $n=1$ and $n_s(\lambda)$.}
    \label{fig:errorTE}
\end{figure}

In order to quantitatively study the differences between different orders of approximation, we compute the percent relative error, $\epsilon$, that we define as
\begin{equation}
    \epsilon=\biggr|1-\frac{\lambda^{asymp}_{n,l}}{\lambda^{exact}_{n,l}}\biggr|\times 100\,.\label{error}
\end{equation}

Fig.~\ref{fig:errorTE} shows $\epsilon$ for TE modes of a $R=5\,\mu$m as a function of $l$ for the first radial order, $n=1$. The relative error follows a similar behavior -although we are now dealing with a nonlinear equation for $\lambda$- as in the previous analyses~\cite{Lam:92,Schiller:93} that consider a constant refractive index: $\epsilon$ decreases as $q$ increases, and it decreases for large values of $l$. Note, however, that $\epsilon$ is not a monotonic function of $l$: $\epsilon$ reaches a maximum for a certain low value of $l^*$, then it decreases up to a new local minimum and, finally, it grows as expected because the asymptotic expansions are in terms of a power of $\nu^{-q/3}$. For completeness, we have included in Fig.~\ref{fig:errorTE} the relative error between the ``exact" $\lambda_{1,l}$ with $n_s=n_s(\lambda)$ and with $n_s=1.45$. In this case, $\epsilon$ reaches a global minimum in $\lambda_{1,l}$ where $n_s(\lambda_{1,l})=1.45$.

\begin{figure}
    \centering    
    \includegraphics[angle=0,width=0.47\textwidth]{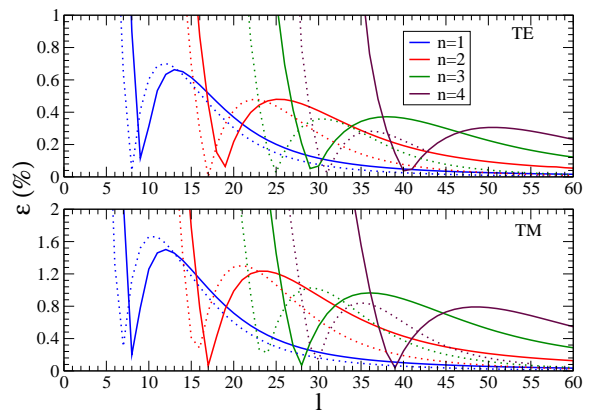}
    \caption{Percent relative error for $R=5\,\mu\mbox{m}$ (solid lines) and $R=1\,\mu\mbox{m}$ (dotted lines), for different values of radial modes ($n=1,2,3,4$), for both TE and TM modes and $n_s(\lambda)$.}
    \label{fig:error}
\end{figure}

Similarly, in Fig.~\ref{fig:error} we show the relative error between the ``exact" and asymptotic ($q=2$) solutions, for both TE and TM modes, for radial modes $n=1,2,3,4$ and for two different values of $R$, namely $R=1\,\mu\mbox{m}$ (dotted lines) and $R=5\,\mu\mbox{m}$ (solid lines), considering the dispersive case $n_s(\lambda)$. 

The behavior of the percent relative error $\epsilon$ does not change considerably for these choices of $n$ and $R$.  

In general, our analysis reveals that the asymptotic expressions for the resonant wavelengths in dispersive media become nonlinear equations once the wavelength-dependent refractive index is taken into consideration. As a consequence, the numerical calculation of $\lambda^p_{n,l}$ becomes as elaborated for the asymptotic expansions as it is for the ``exact" eigenvalue equation, except for the fact that the asymptotic formulas include explicit information on the value of the radial number $n$, simplifying the identification of the associated solutions.
Comparison between the ``exact" and the asymptotic resonant wavelengths reveals convergence in $q$ and, in general, for the particular case of $q=2$, the percent relative error is smaller than $1\%$ for all values of $l>l^*$, for all values of $n$ and for both TE and TM modes. This seemingly small error could be relevant, for example, in high-precision sensing applications~\cite{Hanumegowda:05,Foreman:15} where WGM spectral shifts are of the order of picometers.

\section{Discussion}\label{section4}

\begin{figure} 
    \centering    
    \includegraphics[angle=0,width=0.47\textwidth]{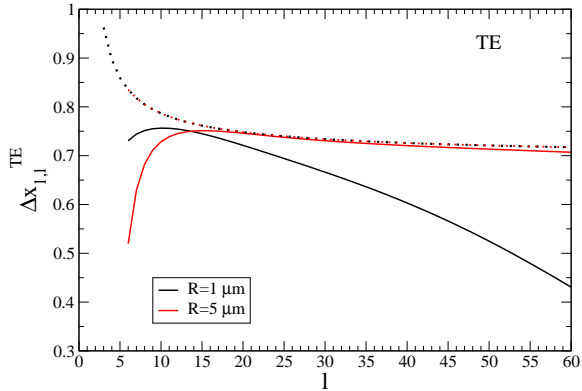}
    \caption{$\Delta x_{1,l}$ computed with $n_s(\lambda)$ (solid lines), and with $n_s~=~1.45$ (dotted lines), for $R=1\,\mu$m and $R=5\,\mu$m and TE modes.}
    \label{fig:FSR}
\end{figure}

Let us now discuss some implications on possible physical observables with the computation of $\lambda_{n,l}^p$ with $n_s(\lambda)$ in comparison with $n_s=$ const.

For instance, the free spectral range (FSR) of a cavity is related to the frequency spacing of two modes with the same values of $n$ but different consecutive azimuthal numbers $l$~\cite{Chiasera:10},
\begin{equation}
\Delta\nu_{n,l}^{p}=c\frac{\Delta x_{n,l}^p}{2\pi R}\,,
\end{equation}
where 
\begin{equation}
\Delta x_{n,l}^p=x_{n,l+1}^p-x_{n,l}^p=2\pi R\left(\frac{1}{\lambda^{p}_{n,l+1}}-\frac{1}{\lambda_{n,l}^p}\right)\,.
\end{equation}

In non-dispersive materials, $\Delta x_{n,l}^p$ is independent on the radius of the sphere and, at first order $(q=1)$, it is given by the inverse of the refractive index: $\Delta x_{n,l}^p~=~1/n_s$.
Nonetheless, resonant WGM in dispersive materials have a nonlinear dependence in $R$, as previously noted and illustrated in Fig.~\ref{fig:frequencies}. In Fig.~\ref{fig:FSR}, we plot $\Delta x_{1,l}^{TE}$ for $R~=~1\,\mu\mbox{m}$ (black lines) and for $R=5\,\mu\mbox{m}$ (red lines). Solid lines are computed for $n_s(\lambda)$ and dotted lines for $n_s=1.45$.  As expected, the plots for non-dispersive media (dotted lines) overlap for different values of $R$; however, differences of the same order ($\mathcal{O}^{(1)}$) appear for dispersive media (solid lines).

On the other hand, in contrast to $\Delta x^p_{n,l}$, we will now show that the $Q$-factor is insensitive to chromatic dispersion.

The intrinsic quality factor $Q$ is an important parameter as it measures the energy losses from a WGM wave propagating in a microsphere resonator. Different physical mechanisms contribute to these losses. For instance, the leakage of the wave through the surface of the sphere is related to radiation losses, $Q_{rad}^{-1}$. Theoretically, this radiation leakage can be computed with knowledge of the imaginary part of the resonant WGM frequency $\omega=2\pi c/\lambda=\omega_R+i\omega_I$. Indeed, $Q_{rad}$ is defined as
\begin{equation}
    Q_{rad}=\frac{1}{2}\frac{\omega_R}{\omega_I}.
    \label{eq:qrad}
\end{equation}

$\omega_I$ has been estimated in~\cite{Datsyuk:92} following a similar expansion as discussed above, but now including the imaginary part of the Hankel and Bessel functions. With this expression for $\omega_I$, $Q_{rad}$ is found to be~\footnote{Observe that this expression includes extra terms and corrects typos presented in~\cite{Datsyuk:92,Buck:03,Talebi:2011}}

\begin{eqnarray}
Q_{rad}&=&\biggr[\frac{\nu}{2}\,n_s^{-(1-2b)}\sqrt{n_s^2(\lambda)-1}\,\times\nonumber\\
&&\left(1-\frac{t_n^0}{2}\left(\frac{\nu}{2}\right)^{-2/3}\right)-\frac{1}{2}\biggr]\,\exp(2T_l),\nonumber\\
\label{qrad}
\end{eqnarray}
\noindent{where}
\begin{equation}
    T_l=\nu(\eta_l-\tanh{\eta_l}),
\end{equation}
\begin{equation}
 b=
    \begin{cases}
   0&\text{for TE modes}\\
1&\text{for TM modes,}
\end{cases}
\label{defb}
\end{equation}
\noindent{and}
\begin{equation}
    \eta_l=\cosh^{-1}\left(\frac{n_s(\lambda)}{1-(\nu)^{-1}\left(t^0_n\left(\frac{\nu}{2}\right)^{1/3}+\frac{n_s(\lambda)}{\sqrt{n_s^2(\lambda)-1}}\right)}\right).
\end{equation}

In addition to $Q_{rad}$, another important energy loss is generated by the absorption of the material. 
This loss contribution is denoted as $Q_{bulk}^{-1}$, and can be approximated by
\cite{Gorodetsky:96}
\begin{equation}
Q_{bulk}=\frac{2\pi n_s(\lambda)}{\alpha(\lambda)\lambda}\,,
\end{equation}
where $\alpha(\lambda)$ is the bulk-fused silica absorption coefficient computed trough Eq.~\eqref{alpha}~\cite{Gorodetsky:00} and shown in Fig.~\ref{fig:index}.

\begin{figure}
    \centering    
    \includegraphics[angle=0,width=0.47\textwidth]{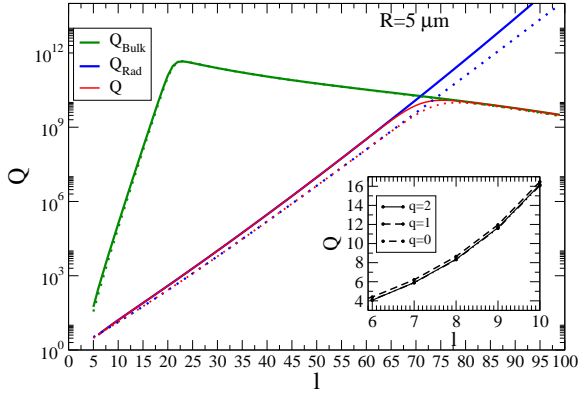}
    \caption{$Q$, $Q_{bulk}$ and $Q_{rad}$ calculated with $\lambda_{1,l}^{exact}$ for TE modes and $R=5\,\mu$m, considering: $n_s(\lambda)$ (solid lines), and $n_s=1.45$ (dotted lines). Inset: $Q$ computed for different asymptotic approximations for low $l$ values.}
    \label{fig:quality}
\end{figure}

\begin{figure}
    \centering    
    \includegraphics[angle=0,width=0.47\textwidth]{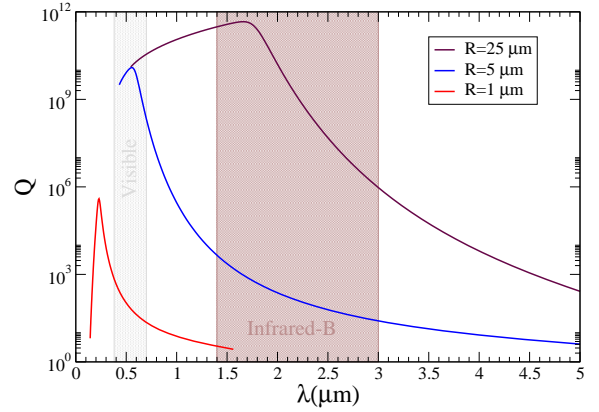}
    \caption{$Q$-factor as a function of the resonant WGM wavelength for different values of the radius of the fused-silica microsphere.}
    \label{fig:Q}
\end{figure}

There are other possible contributions to $Q$ and, in general, the total $Q$-factor (independent of coupling) can be written as 
\begin{equation}
Q^{-1}=Q_{rad}^{-1}+Q_{bulk}^{-1}+\cdots\,.
\end{equation}
Although they can be incorporated, since our purpose here is to discuss the effects of the chromatic dispersion in the $Q$ factor, we neglect them as a first approximation.

Fig.~\ref{fig:quality} shows $Q$ and the contributions $Q_{bulk}$ and $Q_{rad}$ computed for the first radial TE modes of a $R=5\,\mu$m microsphere using $\lambda_{1,l}^{exact}$, for both $n_s(\lambda)$ (solid lines) and $n_s=1.45$ (dotted lines). We can see that, for low $l$ values (corresponding to large $\lambda_{n,l}$), $Q$ is dominated by $Q_{rad}$, while for large $l$ values (corresponding to low $\lambda_{n,l}$),  $Q$ is dominated by $Q_{bulk}$. $Q$ reaches a maximum for a specific value of $\lambda_{n,l}$ that is determined by the interplay between $Q_{rad}$ and $Q_{bulk}$. A comparison between the solid and dotted lines shows that there is no significant effect of the chromatic dispersion. The inset shows $Q$ computed for different asymptotic approximations for low $l$ values, with a similar conclusion.

Finally, as an example of the need of the computation of $\lambda_{n,l}^p$ for designing WGM microspheres for tailored applications, 
the $Q$-factor is estimated as a function of $\lambda$ (see Fig.~\ref{fig:frequencies} to note that there is an univocal relation between $l$ and $\lambda$) for three values of $R$ (see Fig.~\ref{fig:Q}). 
The maximum value of $Q$ corresponds to $l^*$ associated to a specific resonant wavelength. We can see in Fig.~\ref{fig:Q} that, for a microsphere with $R=25\,\mu\mbox{m}$, the maximum $Q$-factor lies in the infrared-B region. This could be relevant because of the recent rising interest in this wavelength band since optical communication systems are approaching their capacity limits. For instance, new technologies allow the transmission of signals in the $2\,\mu\mbox{m}$ band~\cite{Kong:22}. On the other hand, silica microspheres with $R=5\,\mu\mbox{m}$
have a maximum $Q$-factor in the visible range 400\,nm - 700\,nm. 
Applications of WGM modes are very diverse. Even low-$Q$ microspheres could have potential applications, for instance, in sensing applications in the high loss regime~\cite{Weller:08}, or in the enhancement of the energy efficiency of solar-cells~\cite{Grandidier:11,Yao:12}. 
Our estimation of the $Q$-factor shows that a $R=1\,\mu\mbox{m}$ silica microsphere will have a 
low-$Q$ in the visible spectrum, suggesting an optimal size to enhance the energy efficiency of solar-cells. 

\section{Conclusions}\label{conclusions}

Motivated by the need to study different morphologies and materials to aid in the design and construction of WGM microresonators for tailored applications, we have presented an analysis of the effect of chromatic dispersion in the computation of the WGM resonant wavelengths. We have obtained the solutions to the characteristic equation~\eqref{eqcar} and the asymptotic approximations~\eqref{Eq:Lam}. We have verified the accuracy of the asymptotic formulas in dispersive materials, $n_s(\lambda)$, and found that it is a good approximation with an error around $0.1 \%$ for $l \gg 1$. This is valid for different radii and different radial modes. 
There are, however, larger differences of the order of a few percent) for low values of $l$, where the wavelengths are of the order of $R/2\pi$, for all $n$ (See Fig.~\ref{fig:error}). 
Although there are changes in the WGM resonant wavelengths, we have noticed that there are no significant changes in the radial field distribution, even for the low order cases.
More importantly, because of the nonlinearity introduced by the chromatic dispersion, there is no rescaling in the resonant WGM wavelengths with the radii of the sphere. This is an important difference in comparison with the non-dispersive case. By computing the distance between two resonant size parameters, $\Delta x_{n,l}^p$, we have found that this dependence in the radius of the sphere could be relevant in the extraction of physical parameters that are important in the implementation of these microresonators. Additionally, a corrected expression of the radiative contribution to the $Q$-factor was found. Using this expression, we have computed $Q$ including the radiative and the bulk contributions and found that it is insensitive to the effect of chromatic dispersion. Nevertheless, $Q_{rad}$ changes significantly for different radius and thus, different sizes of the microspheres could have different applications depending on the desired operating wavelength.
Therefore, our study presents a pathway for designing microsphere resonators suitable for various applications spanning a wide range of wavelengths.

\bmhead{Acknowledgements}
This work was partially supported by CONAHCYT-SNII.

\bibliographystyle{unsrt}
\bibliography{biblio}

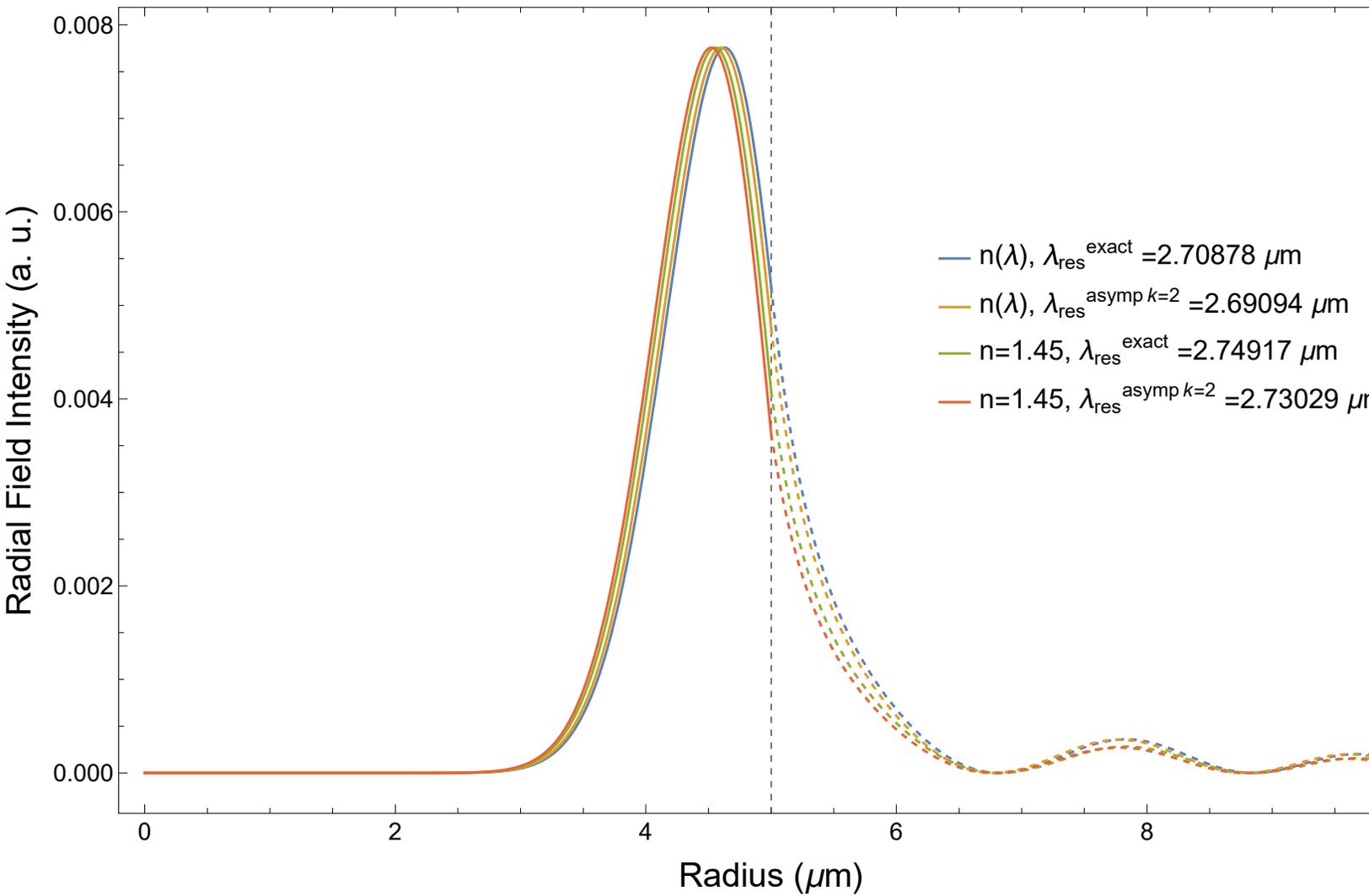
\end{document}